\documentclass[aps,floats,epsf,twocolumn]{revtex4}

\begin{document}

\title{Time reversal, fermion doubling, and the masses of lattice
Dirac fermions in three dimensions}

\author{Igor F. Herbut}

\affiliation{Department of Physics, Simon Fraser University,
 Burnaby, British Columbia, Canada V5A 1S6}

\begin{abstract} Motivated by recent examples of three-dimensional lattice Hamiltonians with
massless Dirac fermions in their (bulk) spectrum, I revisit the problem of fermion doubling on bipartite lattices. The
number  of components of the Dirac fermion in a time-reversal and parity invariant d-dimensional lattice system is determined by the minimal
representation of the Clifford algebra of $d+1$ Hermitian Dirac matrices that allows a construction of the
time-reversal operator with the square of unity, and it equals $2^d$ for $d=2,3$. Possible mass-terms for (spinless) Dirac  fermions
are listed and discussed. In three dimensions there are altogether eight independent masses, out of which four are even, and four are
odd under time reversal. A specific violation of time-reversal symmetry that leads to (minimal) four-component massless Dirac fermion in three
dimensions at low energies is constructed.

\end{abstract}
\maketitle

\vspace{10pt}

\section{Introduction}

It is well-known that the tight-binding Hamiltonian for spinless fermions on a honeycomb lattice exhibits two inequivalent zero-energy points in the
  Brillouin zone,  in vicinity of which the excitation energy has an approximate linear momentum dependence \cite{semenoff}. Honeycomb lattice being bipartite,
  this implies that the low-energy excitations behave like four-component Dirac fermions. Another well-known two-dimensional example of
  four-component Dirac fermions arising on a square lattice is the $\pi$-flux Hamiltonian \cite{affleck}, in which there is a single Dirac point, but the unit cell consists of four lattice sites. Of course, this is precisely how it has to be according to the Nielsen-Ninomiya (NN) theorem\cite{nielsen}, which  forbids the appearance of two-component (Weyl) fermions on a lattice under a certain set of conditions, which, importantly, includes the time-reversal invariance of the Hamiltonian.

 Several ways of circumventing the NN ``no-go" theorem do exist, however. It was pointed out \cite{fradkin} that it may be possible to find a two-component massless fermion on a two-dimensional lattice at the expense of having additional low-energy excitations in the system, such as flat bands, for example. Furthermore, breaking the time-reversal and then fine-tuning the hopping on the honeycomb lattice can also lead to a single Weyl fermion at low-energies\cite{haldane}. Since neutrinos are such Weyl fermions, a possibility of their lattice realization has over the years attracted a considerable attention in the lattice field theory community \cite{rothe}. In condensed matter physics Weyl fermions would represent a new class of quasiparticle excitations with interesting properties, presently under scrutiny in the context of topological band insulators, where they appear generically as surface modes.\cite{hasan}

  Recently, interesting generalizations of both $\pi$-flux and graphene's lattice Hamiltonians  to three dimensions (3D) have appeared \cite{hosur}, both displaying (for spinless fermions) eight-component Dirac fermions as their low-energy excitations. Since the number of order parameters, or the mass terms, in 3D is larger than in 2D, it seems there could be more possibilities for a reduction of the dimension of the low-energy Dirac fermion by fine tuning the masses {\it a la Haldane}. This issue is investigated here, to find that generally the ``smallest" Dirac fermion that can be found in the spectrum by breaking the time-reversal in 3D still has four components. Along the way, it is shown that the minimal  dimension of $2^d$ of the Dirac fermion in a d-dimensional time-reversal-symmetric lattice Hamiltonian can also be understood as a purely algebraic constraint: a smaller representation of Dirac matrices does not allow a construction of a time-reversal operator for the Dirac fermion that would square to $+1$, as it must for the spinless fermions hopping on a lattice.

  A $2^d$-component low-energy massless Dirac Hamiltonian in d-dimensions exhibits $O(d+1)$ ``chiral" symmetry that may be used to classify possible (order parameter) mass terms. In a representation-independent way, I show that in general there are $d+1$ time-reversal invariant mass terms, which form a vector under the chiral rotations. In addition, there are also one (in $d=2$) and four (in $d=3$) masses that break time-reversal invariance, and which are  scalar and vector under the chiral symmetry, respectively.  A possible physical meaning of these masses on the 2D honeycomb \cite{semenoff, haldane, khveshchenko, hou, hjr} and the 2D \cite{tanaka, malcolm} and 3D \cite{hosur} $\pi$-flux lattices has been discussed at length elsewhere.  The commutation relations between the mass terms in 3D allow them to be fine-tuned so as to reduce the number of components of the low-energy Dirac excitation, albeit only to four.

\section{Fermion doubling: algebraic derivation}

Assume that {\it all} low-energy excitations of a  lattice Hamiltonian for spinless electrons hopping in $d$ spatial dimensions  may be described by an effective Dirac Hamiltonian
\begin{equation}
H= \sum_{i=1}^{d} \alpha_i p_i,
\end{equation}
where $p_i$ are the usual momentum operators, $\alpha_i$ are Hermitian Dirac matrices satisfying the Clifford algebra
$\{ \alpha_i , \alpha_j \} = 2 \delta_{ij}$, of yet to be determined dimensionality. The Fermi velocity in the above expression is set to unity.

  Let us deduce the dimension of $\alpha$-matrices in the effective Dirac Hamiltonian in 3D ($d=3$), under two assumptions. First, assume that inversion symmetry is obeyed by the underlying lattice  Hamiltonian. Since it then also needs to be respected within the low-energy Dirac sector, it means that there exists a fourth, {\it parity}  matrix, $\beta$, such that
  \begin{equation}
  \{ \beta, \alpha_i \}=0.
  \end{equation}
  Note that this is automatically the case for the bipartite lattice Hamiltonians. Second, let us assume the time-reversal symmetry. Since  the lattice Hamiltonian describes the real-space motion of spinless particles, in the basis of lattice sites the antilinear operator $I_t$ for the time-reversal is simply that of complex conjugation, which then satisfies $I_t ^2 = 1$. Since, on the other hand, the value of $I_t^2$  is basis independent\cite{remark}, it is the same in the basis of Bloch states. This means that for a time-reversal lattice Hamiltonian there must exist an operator $I_t$ in the low-energy Dirac sector which squares to unity. I show next that this immediately implies that the Dirac Hamiltonian must be at least eight-dimensional, i. e. the usual ``fermion doubling".

  The mathematical basis behind the above assertion is the Okubo's theory of real representation of Clifford algebras \cite{okubo}. Here I reproduce only the part relevant for the present purposes. Define a Clifford algebra $C(p,q)$ of $p+q$ mutually anticommuting objects, $ p $ ($q$) of which square to $+1$ ($-1$). The smallest purely real representation of $C(p,q)$ is then $2^q$-dimensional, and it exists when $p=q$ or $p=q+1$ (mod 8). Okubo's theory, of course, contains a great deal more than this corollary. An important implication of this result is that a $2^n$-dimensional complex representation of the Clifford algebra $ C (2n+1,0)$, where all elements square to $+1$, as in the Dirac Hamiltonian in Eq. 1, can always be chosen so that $n$ matrices are imaginary, and the remaining $n+1$ real \cite{herbut}. It is easy to check that this is indeed the case for $n=1$ and $n=2$, for example, by a direct construction.

    Returning to the 3D Dirac Hamiltonian, let us construct the  time-reversal operator for the low-energy excitations.
    The matrices $\{ \alpha_1, \alpha_2, \alpha_3, \beta \} $ form
    the Clifford algebra $C(4,0)$, and the minimal complex representation, as well-known, is four-dimensional. We can form maximally five mutually anticommuting matrices by constructing the fifth (Hermitian) matrix $\alpha_1\alpha_2 \alpha_3 \beta$, which together with the first four then forms the Clifford algebra $C(5,0)$. Okubo's  general theory, or a simple inspection in this case, implies then that out of these five four-dimensional matrices we can have at most two as purely imaginary. So let us chose $\alpha_1$ and $\beta$ real, and $\alpha_2$ and $\alpha_3$ imaginary, for example. All other choices are, of course, related to this one by a unitary transformation.  The time-reversal operator is $I_t = U K$, where $U$ is unitary and $K$ stands for complex conjugation. It then follows that for our choice the unitary part must satisfy
    \begin{equation}
    \{ U,\alpha_1 \} = [ U,\alpha_2 ]  = [ U, \alpha_3 ]  =0,
    \end{equation}
    which has two solutions: $U= e^{i\phi} \alpha_1 \beta $, or $U= e^{i\phi} \alpha_2 \alpha_3 \beta$, with $\phi$ arbitrary. Demanding that a spatial symmetry such as parity commutes with the time-reversal in the low-energy Dirac sector, as it does in the microscopic lattice Hamiltonian, selects
     the latter solution. Irrespectively of this condition it follows that
   \begin{equation}
    I_t ^2 = -1,
    \end{equation}
    which stands in contradiction to the lattice origin of the Dirac Hamiltonian. One concludes that a four-dimensional Dirac fermion cannot represent all the low-energy excitation of a time and space-reversal invariant Hamiltonian on a 3D lattice.

    A (reducible) eight-component representation however, is compatible with the time-reversal symmetry: we can choose all three $\alpha$-matrices to be imaginary, for example, so that the unitary operator $U$ commutes with all three. This implies that $U= e^{i\phi}$, or $U= e^{i\phi} \alpha_1 \alpha_2 \alpha_3$. In either case it is now $I_t ^2 = +1$. Commutation of the parity matrix $\beta$ and the time-reversal in the low-energy Dirac sector implies further that  $U=e^{i\phi}$. Choosing the phase $\phi =0$ then leaves a simple $I_t = K$.

    Similar reasoning can straightforwardly be applied to $d=2$ as well. Assuming $\alpha_1$, $\alpha_2$ and $\beta$ to be two-dimensional matrices, implies that one of them can be chosen imaginary, and the other two real. Choosing $\beta$ imaginary makes it obvious that the time-reversal operator that would also commute with parity does not even exist. Going to a four-dimensional representation, on the other hand, allows one to chose $\alpha_1$ and $\alpha_2$ imaginary and $\beta$ real, so that time-reversal operator becomes complex conjugation, the same as in 3D.

\section{Masses and symmetries}

   Let us now consider possible mass-terms that upon addition to the Dirac Hamiltonian in Eq. 1 would gap its spectrum. On a lattice such masses would break some of the symmetry and represent certain order parameter. Before considering three dimensions, let me review the familiar example of 2D first.

   \subsection{ Two dimensions}

   If $d=2$, we can chose the two four-dimensional matrices $\alpha_1$ and $\alpha_2$ to be imaginary, as discussed above. This leaves three other linearly independent real Hermitian four-dimensional matrices $\{ \beta_1, \beta_2, \beta_3 \}$, which together with $\alpha$-matrices form the maximal Clifford algebra $C(5,0)$ of dimension four. One of the $\beta$-matrices may be chosen to represent parity operator, and of course, $\beta_3= \alpha_1 \alpha_2 \beta_1 \beta_2$.  The time-reversal operator is then just the complex conjugation, and the most general time-reversal invariant mass-term is $\vec{m} \cdot \vec{V}_R$, where
   \begin{equation}
   \vec{V}_R = (\beta_1, \beta_2, \beta_3 )
   \end{equation}
   is a vector under the $O(3)$ chiral symmetry of the Dirac Hamiltonian, generated by $i\beta_i \beta_j$, for $i\neq j$. The masses $\vec{m}=(m_1,m_2,m_3)$ are  real.

Multiplying an odd number of $\beta$-matrices produces another matrix that anticommutes with $\alpha_1$ and $\alpha_2$, and thus leads to a new linearly independent mass term. For $d=2$ there is only one such product,
\begin{equation}
S_I = i \beta_1\beta_2\beta_3,
\end{equation}
where the imaginary unit is necessary to make the matrix Hermitian. This mass is evidently a scalar under $O(3)$ symmetry, but {\it odd} under time-reversal. On the honeycomb lattice $\vec{V}_R$ represents a charge density wave, and two components of a Kekule bond density wave \cite{hjr}. $S_I$, on the other hand, represents the anomalous quantum Hall state\cite{haldane}.

  \subsection{Three dimensions}

  In $d=3$, we chose again the three eight-dimensional Hermitian matrices $\alpha_i$, $i=1,2,3$ to be imaginary, to have simply $I_t = K$. Now, however, there are two, related but distinct ways to complete this set to the maximal number of seven mutually anticommuting matrices. First, we can chose four real matrices
  $\beta_i$, $i=1,2,3,4$, which together with $\alpha$-matrices form $C(7,0)$. Any linear combination  $\vec{m} \cdot \vec{V}_R$ of these would yield a time-reversal invariant mass term, with
  \begin{equation}
   \vec{V}_R = (\beta_1, \beta_2, \beta_3, \beta_4 ),
   \end{equation}
again as a real vector under $O(4)$ symmetry of the Dirac Hamiltonian. This is analogous to the situation in $d=2$. The novelty in $d=3$ is that there are now four distinct three-products of $\beta$-matrices, which are also possible mass terms. Consider a $O(4)$ vector
 \begin{equation}
   \vec{V}_I = (\tilde{\beta}_1,\tilde{\beta}_2,\tilde{\beta}_3,\tilde{\beta}_4) ,
   \end{equation}
 with $\tilde{\beta}_i = i \beta_i \beta_1\beta_2 \beta_3\beta_4 $. $\vec{V}_I$ is clearly odd under time-reversal, due to the imaginary unit necessary to render the matrices Hermitian.

 A concrete realization of the above algebra on a cubic lattice with a $\pi$-flux through a plaquette was recently proposed and discussed by Hosur, Ryu, and Vishwanath  \cite{hosur}. Time-reversal-even matrices in their example represent the charge-density and bond-density wave order parameters, and the time-reversal-odd order parameters are three anomalous Hall states and a ``chiral" topological insulator.

\section{Four-component massless fermion in 3D: a construction}

   Finally, let us now investigate the possibility of having a Dirac massless fermion with fewer than eight components in the spectrum by a judicious breaking of time-reversal symmetry. I consider only $d=3$ in detail, although the result for $d=2$, which has been pointed out by Haldane long ago, would be analogous.

   In $d=3$ the $\beta$-matrices obey the commutation relations:
   \begin{equation}
   \{ \beta_i, \beta_j \} = \{ \tilde{\beta}_i, \tilde{\beta}_j \} = 2\delta_{ij},
   \end{equation}
   \begin{equation}
   \{ \beta_i, \tilde{\beta}_i \} = [ \beta_i, \tilde{\beta}_{j\neq i} ] =0.
   \end{equation}
   Consider then the most general massive Dirac Hamiltonian in d=3:
   \begin{equation}
 H= \sum_{i=1} ^3 \alpha_i p_i + \vec{m} \cdot \vec{V}_R + \vec{ \tilde{m} } \cdot \vec{V}_I.
   \end{equation}
   Without loss of generality we can utilize the  $O(4)$ chiral invariance  and chose $\vec{V}_R = (\beta_1, 0,0,0)$, so that
   $\vec{m} = (m,0,0,0)$. Squaring the Dirac Hamiltonian we then find
   \begin{equation}
   H^2 = p^2 + m^2 + \tilde{m}^2 + m \sum_{i=2}^4 \tilde{m}_i \{ \beta_1,\tilde{\beta}_i \}
   \end{equation}
   where $\tilde{m}^2 = \sum_{i=1}^4 \tilde{m}_i ^2 $.  Consider the three matrices
   \begin{equation}
   \gamma_i = \frac{1}{2} \{ \beta_1,\tilde{\beta}_i \} = \beta_1 \tilde{\beta}_i,
   \end{equation}
   $i=2,3,4$, that appear in the last expression. Evidently,
   \begin{equation}
   \{ \gamma_i, \gamma_j \} = 2\delta_{ij},
   \end{equation}
   and they themselves close the Clifford algebra $C(3,0)$.   Since the two inequivalent irreducible representations of $C(3,0)$ are two dimensional,
   the three $\gamma$-matrices can be brought into a block-diagonal form, with the blocks as Pauli matrices. This implies that there exists a unitary transformation that puts the (square of) Hamiltonian into the form
   \begin{equation}
   H^2 = p^2 + m^2 + \tilde{m}^2 + 2 m (\tilde{m}_2 ^2 + \tilde{m}_3 ^2 + \tilde{m}_4 ^2 )^{1/2} ( I_4 \otimes \sigma_3 ) ,
   \end{equation}
  where $I_4$ is a four-dimensional unit matrix. The energy spectrum is therefore
  \begin{equation}
  E_\pm ^2 = p^2 + (m \pm   ( \tilde{m}_2 ^2 + \tilde{m}_3 ^2 + \tilde{m}_4 ^2 )^{1/2}  )^2 + \tilde{m}_1 ^2,
  \end{equation}
  with each sign being four times degenerate.

    Since choosing the mass $\vec{V}_R$ breaks the $O(4)$ symmetry, the obtained spectrum depends  qualitatively on the ``direction"
 of the vector $\vec{V}_I$. Choosing $\vec{V}_I$ parallel to $\vec{V}_R$ adds the corresponding masses in squares and only increases the
 mass gap.  Taking $\vec{V}_I$ to be orthogonal to $\vec{V}_R$, on the other hand, increases the mass gap for one half, and decreases it for the other half
 of the components of the original eight-component Dirac fermion. In particular, for $\tilde{m}_1 =0$ and
 \begin{equation}
 m^2 = \tilde{m}_2 ^2 + \tilde{m}_3 ^2 + \tilde{m}_4 ^2
 \end{equation}
 the mass-gap for four of the components collapses to zero, and one finds a four-component massless Dirac fermion as a sole
 low-energy excitation. It is obvious from the above derivation, however, that no combination of eight mass-terms could leave a two-component
 Weyl fermion at low energies in $d=3$. This is possible only in $d=2$, where the Weyl fermion results from halving the original four-component  Dirac fermion\cite{haldane}, in precise analogy to the above derivation in 3D.

 \section{Conclusion and discussion}

 I studied massless Dirac fermions as the low-energy excitations of lattice Hamiltonians in two and three dimensions in a general,
 representation-independent way. The existence of the time-reversal operator with a trivial square in the low-energy
sector is shown to imply that the Dirac fermion must have at least four and eight components, in 2D and 3D, respectively.
In 3D there are eight possible mass terms for (spinless) Dirac fermions, four of which violate time-reversal symmetry. A combination
of time-reversal-even and time-reversal-odd masses that leaves a four-component massless Dirac fermion in the spectrum in 3D was constructed.

  Recently, an interesting variation of the $\pi$-flux Hamiltonian in 2D was proposed \cite{malcolm} which seemingly has two Weyl fermions with {\sl different}
  velocities as its low-energy spectrum, although it is perfectly time-reversal and parity invariant.
  It is interesting to see why this {\it does not} contradict our assertion that the minimal dimension
  of a Dirac fermion in $d$ dimensions is $2^d$. The Hamiltonian of the ref. 13 can be written as
  \begin{equation}
  H= (\alpha_1 + \delta i \beta_1 \beta_2) p_1 + (\alpha_2 + \delta i \beta_1 \beta_3) p_2
  \end{equation}
  in the notation of sec. III a, with $\delta$ as a real parameter.
  The energy spectrum is then $\pm (1\pm \delta) p$, and choosing $\alpha_i$ imaginary and $\beta_i$ real, the time-reversal
  operator which commutes with the Hamiltonian is $I_t=K$, in accord with the lattice origin of the Hamiltonian. The contradiction with our assertion is avoided by the fact that $H$ {\sl cannot} be transformed into a block-diagonal form that would describe an orthogonal sum of two Weyl Hamiltonians.
   This can be proven by {\sl reductio ad absurdum}. If it could be so block-diagonalized, it would have to be equivalent to
  \begin{equation}
  H_d = (1+\delta)(p_1 \sigma_1 + p_2 \sigma_2 ) \oplus  (1-\delta) (p_1 \sigma_1 + p_2 \sigma_2 ),
  \end{equation}
  and therefore there would exist a traceless matrix that squares to unity, and that commutes with $H$ for any momentum, which would be equivalent to $\sigma_3 \otimes \sigma_0$. But, such a matrix obviously does not exist: the only matrix that commutes with $\alpha_1$, $\alpha_2$, and $i\beta_1 \beta_2$ is $i\beta_1 \beta_2$ itself, which however fails to commute with $i\beta_1\beta_3$. The existence of matrix $\beta_1$ which anticommutes with the Hamiltonian in Eq. 18, on the other hand, implies that it can be transformed into a block off-diagonal form. This is a nice example of how the presence of additional degrees of freedom at arbitrary low energy only seems to invalidate the NN theorem. Note that putting the parameter $\delta=1$ in the above Hamiltonian makes two of the bands flat and with zero energy, somewhat similar to the early example of Dagotto, Fradkin, and Moreo \cite{fradkin}.

Several extensions of the present work can be envisaged. One can include the particle's spin and/or consider off-diagonal pairing terms in 3D to determine all possible Dirac mass terms, similarly to what has been done in 2D \cite{chamon}. Extensions to higher dimensions, albeit somewhat academic from the point of view of a condensed matter physicist, may be interesting from a particle physics or a methodological perspective. Finally, it could be interesting to investigate the structure of topological defects in the mass-terms in 3D, as several analogies with the more familiar defects in 2D \cite{herbut1} are known to exist. \cite{teo, lu}

\section{Acknowledgement}

 This work is supported by the NSERC of Canada. The author is grateful to Chi-Ken Lu for many useful discussions.


\begin{thebibliography}{99}
\bibitem{semenoff} G. W. Semenoff, Phys. Rev. Lett. {\bf 53}, 2449 (1984).
\bibitem{affleck} I. Affleck and J. B. Marston, Phys. Rev. B {\bf 37}, 3774 (1988).
\bibitem{nielsen} H. B. Nielsen and M. Ninomiya, Nucl. Phys. {\bf 185}, 20 (1981).
\bibitem{fradkin} E. Dagotto, E. Fradkin, A. Moreo, Phys. Lett. {\bf 172}, 383 (1986).
\bibitem{haldane} F. D. M. Haldane, Phys. Rev. Lett. {\bf 61}, 2015 (1988).
\bibitem{rothe} See for example, H. Rothe, {\sl Lattice Gauge Theories, An Introduction}, (World Scientific, 2005).
\bibitem{hasan} M. Z. Hasan and C. L. Kane, Rev. Mod. Phys. {\bf  82}, 3045  (2010).
\bibitem{hosur} P. Hosur, S. Ryu, and A. Vishwanath, Phys. Rev. B {\bf 81}, 045120 (2010).
\bibitem{khveshchenko} D. V. Khveshchenko, Phys. Rev. Lett. {\bf 87}, 246802 (2001)
\bibitem{hou} C-Y. Hou, C. Chamon, and C. Mudry, Phys. Rev. Lett. {\bf 98}, 186809 (2007).
\bibitem{hjr} For a review, see I. F. Herbut, V. Juri\v ci\' c, and B. Roy, Phys. Rev. B {\bf 79}, 085116 (2009).
\bibitem{tanaka} A. Tanaka and and X. Hu, Phys. Rev. Lett. {\bf 95}, 036402 (2005).
\bibitem{malcolm} M. P. Kennett, N. Komeilizadeh, K. Kaveh, P. M. Smith, Phys. Rev. A {\bf 83}, 053636 (2011).
\bibitem{remark} This assertion is easy to prove. See, for example, K. Gotfried and T-M. Yan, {\sl Quantum Mechanics: Fundamentals},
2nd ed., (Springer, 2004), Section 7.2.
\bibitem{okubo} S. Okubo, J. Math. Phys. {\bf 32},  1657 (1991).
\bibitem{herbut} I. F. Herbut and C.-K. Lu, Phys. Rev. B {\bf 82}, 125402 (2010).
\bibitem{chamon} S. Ryu, C. Mudry, C-Y. Hou, C. Chamon, Phys. Rev. B {\bf 80}, 205319 (2009).
\bibitem{herbut1} I. F. Herbut, Phys. Rev. Lett. {\bf 99}, 206404 (2007); {\sl ibid.} {\bf 104},  066404 (2010).
\bibitem{teo} J. C. Y. Teo and C. L. Kane, Phys. Rev. Lett. {\bf 104}, 046401 (2010).
\bibitem{lu} I. F. Herbut and C.-K. Lu, Phys. Rev. B {\bf 83}, 125412 (2011).

\end{thebibliography}
\end{document}